\documentclass[runningheads]{llncs}

\usepackage[utf8x]{inputenc}
\usepackage[T1]{fontenc}
\usepackage[normalem]{ulem}
\usepackage{amsmath}
\usepackage{amssymb}
\usepackage{epigraph}
\usepackage{hyperref}
\usepackage{mathtools}
\usepackage{color}
\definecolor{keywordcolor}{rgb}{0.7, 0.1, 0.1}   % red
\definecolor{tacticcolor}{rgb}{0.1, 0.2, 0.6}    % blue
\definecolor{commentcolor}{rgb}{0.4, 0.4, 0.4}   % grey
\definecolor{symbolcolor}{rgb}{0.0, 0.1, 0.6}    % blue
\definecolor{sortcolor}{rgb}{0.1, 0.5, 0.1}      % green
\definecolor{attributecolor}{rgb}{0.7, 0.1, 0.1} % red
\usepackage{xspace}
\usepackage{tikz}
\usepackage{ellipsis}
\usepackage[stable]{footmisc}

\newcommand{\mmO}{\textsf{MM0}}
\newcommand{\fresh}{\mathrel{\#}}
\newcommand{\wff}{\mathsf{wff}\ }
\newcommand{\ctx}{\ \mathsf{ctx}}
\newcommand{\ok}{\ \mathsf{ok}}
\newcommand{\sort}{\,\mathsf{sort}}

\newcommand{\env}{\ \mathsf{env}}

\newcommand{\alloc}{\mathrm{alloc}}
\newcommand{\hto}{\hookrightarrow}
\newcommand{\uto}{\hookrightarrow_{\mathsf{u}}}
\newcommand{\coconv}{\mathrel{\smash{\overset{?}{\equiv}}}}
\begin{document}
%
%% Title information
\title{Metamath Zero: The Cartesian Theorem Prover}
\titlerunning{Metamath Zero}
%
%% Author information
\author{Mario Carneiro\orcidID{0000-0002-0470-5249}}
\authorrunning{M. Carneiro}
\institute{Carnegie Mellon University, Pittsburgh PA\\
\email{mcarneir@andrew.cmu.edu}}
\maketitle              % typeset the header of the contribution
\begin{abstract}
As the usage of theorem prover technology expands, so too does the reliance on correctness of the tools. Metamath Zero is a verification system that aims for simplicity of logic and implementation, without compromising on efficiency of verification. It is formally specified in its own language, and supports a number of translations to and from other proof languages. This paper describes the abstract logic of Metamath Zero, essentially a multi-sorted first order logic, as well as the binary proof format and the way in which it can ensure essentially linear time verification while still being concise and efficient at scale. Metamath Zero currently holds the record for fastest verification of the \texttt{set.mm} Metamath library of proofs in ZFC (including 71 of Wiedijk's 100 formalization targets), at less than 200 ms. Ultimately, we intend to use it to verify the correctness of the implementation of the verifier down to binary executable, so it can be used as a root of trust for more complex proof systems.

\end{abstract}
\section{Introduction}\label{sec:intro}

The idea of using computers to check mathematical statements has been around almost as long as computers themselves, but the scope of formalizations have grown in recent times, both in pure mathematics and software verification, and it now seems that there is nothing that is really beyond our reach if we aim for it. But at the same time, software faces a crisis of correctness, where more powerful systems lead to more reliance on computers and higher stakes for failure. Software verification stands poised to solve this problem, providing a high level of certainty in correctness for critical components.

But software verification systems are themselves critical components, particularly the popular and effective ones. A proof in such a system is only as good as the software that checks it. How can we bootstrap trust in our systems?

This paper presents a formal system, called Metamath Zero (\mmO), which aims to fill this gap, having both a simple extensible logical theory and a straightforward yet efficient proof format. Work to prove the correctness theorem is ongoing, but this paper explains the design of the system and how it relates to other theorem provers, as well as general considerations for any bootstrapping theorem prover.

\subsection{Who verifies the verifiers?}

There are two major sources of untrustworthiness in a verification system: the logic and the implementation. If the logic is unsound, then it may be able to prove absurd statements. This problem is well studied in the mathematical and logical literature, and there are a number of formal systems that are widely believed to be trustworthy, such as Peano Arithmetic (\textsf{PA}) and Zermelo-Fraenkel set theory (\textsf{ZFC}), and moreover the relationship between these theories and others (such as type theory, higher order logic, etc.) are well understood.

Much more concerning is implementation correctness. Implementation bugs can exist in the theorem prover itself, the compiler for the language, any additional components used by the compiler (the preprocessor, linker, and assembler, if applicable), as well as the operating system, firmware, and hardware. In this area, mathematics and logic holds little sway, and it is ``common knowledge'' that no nontrivial program is or can be bug-free. The argument for correctness of these systems is largely a social one: the compiler has compiled many programs without any bugs (that we noticed) (except when we noticed and fixed the bugs), so it must work well enough.

What can we do? Our strategy is to start from the ground up, defining all the properties that the verifier system of our dreams should have, and then start building it. With the right set-up, the goal turns out to be surprisingly achievable. Here are some dream goals:

\begin{enumerate}
  \item It should be proven correct down to the lowest possible level. Some options for the lowest level include:
  \begin{enumerate}
    \item a logical rendering of the code;
    \item the code itself, inside a logical rendering of the language;
    \item the machine code, specified given an ISA (instruction set architecture);
    \item the computer, down to the logic gates that make it up;
    \item the fabrication process relative to some electrical or physical model.
  \end{enumerate}
  \item It should permit the user to prove any theorem they like.
  \item It should permit the user to write any program they like and prove any theorem about that program.
  \item There should be no practical upper limit on the complexity of the target programs, and they should be able to run as fast as the machine is capable.
  \item It should be fast.
  \item It should be easy to use.
\end{enumerate}

While there is no theoretical reason not to push (1) all the way to level (e), the drawback of the most aggressive levels (d) and (e) is that it limits redistribution capabilities. If you prove a system correct at level (e), the proof only holds if you use the given fabrication process, and similarly (d) only holds if you use the given chip design. A proof relative to (c) holds as long as the ``reader'' has the same ISA as the author, so if we pick a relatively popular ISA then we can write proofs that are independently verifiable by a large segment of the population. Furthermore, in order to prove a chip design correct, it must be accessible at least to the proof author, and independent verification of the proof also essentially entails exposing the chip design, which excludes closed source chip designs, which are by far the majority in consumer hardware. For these reasons, \mmO\ targets (c) with the Intel x86-64 architecture on Linux.

There are open source ISAs available; the most prominent as of this writing is RISC-V. However, while we hope to extend the \mmO\ bootstrap to encompass alternative architectures in future work, we only need a freely available ISA for the purpose of proof, and beyond that the most important factor for independent verification is general availability. However, we recognize that these are contingent facts, and different choice may be better in the future.

To satisfy (2), \mmO\ is a logical framework, with ``pluggable axioms'' to support any desired mathematical foundation. To satisfy (3) and (4), we implemented, in \mmO, a specification for x86-64, so that users can write any program they like. To satisfy (5), proofs will be laid out such that checking them is as straightforward as possible, roughly linear time.

(6) is a subjective criterion, and also one we are willing to compromise on in favor of the others. Nevertheless some degree of ease of use is needed in order to get enough done to achieve the other goals. To that end, \mmO\ the verifier has a front end \textsf{MM1}, which provides some ITP (interactive theorem prover) features, in an \emph{unverified} setting. This is an extension of the LCF-style prover architecture to completely separate and isolate the verifier from the prover.

In particular, a compiler for producing verified programs would sit at this unverified prover level, producing machine code from high level code so that users don't have to write machine code. This compromises (4) to some extent if the compiler produces poor code, but that has an obvious mitigation.

\subsection{Efficiency matters}

Why should it matter if a proof takes hours or days to compile? Besides the obvious problem that no one likes to set up a build job that takes hours, a longer-running proof means a larger window for ``attack'' from the outside world: more cosmic rays, more OS context switches, more firmware updates, more hardware failures. Generally speaking, getting your work done faster means less possibility of interference of all kinds.

But these are usually negligible concerns. Most of the time a bug either manifests immediately or not at all during a run. An exception is out of memory handling bugs, which are more likely to be exercised in a memory intensive process, so using less memory is one way to mitigate (but not avoid) this problem. Memory allocation bugs are distressingly common, because allocation failure is so rare that the error pathways are almost never tested.

But sometimes, performance is about more than just getting work done a little faster. When something takes \emph{a lot} less time, it changes the way you interact with the computer. A process that takes hours goes on the nightly build server; a process that takes minutes might be a compile that runs on your local machine; a process that takes seconds is a progress bar; and a process that takes milliseconds might happen in an editor between keystrokes. A program that can verify proofs of OS kernel correctness in under a second can work in the boot stage, providing ``secure boot'' backed not by social factors but by mathematical proof.

Furthermore, a verifier is a \emph{component} in a larger system. What is reasonable performance for an end-user tool may not be reasonable for a backend library. Program correctness proofs are generally large, so the thing that processes the proof needs to be fast. (The proof authoring tool should also be fast, but this work can be cached and broken down when the expense becomes too much, while verification cannot be cached without adding the caching process to the trusted part.) As proofs become larger, the performance of the kernel will inevitably become the final bottleneck, while on the other hand a ``nightly build'' can't be nightly if it takes more than 24 hours, so improving kernel performance ultimately translates to scaling the size of checkable proofs.

The \texttt{set.mm} Metamath library of theorems in \textsf{ZFC} contains over 34000 proofs, and \mmO\ can check a translation of it in $195\pm5$ ms. This can be (unfairly) compared to libraries in other systems such as Isabelle, Coq, or Lean that take over an hour to compile their standard library and sometimes much more, a difference of 4--5 orders of magnitude. While this is not definitive evidence, such a vast discrepancy indicates that architectural differences matter, and significant gains are possible over the status quo.

\subsection{A standalone verifier}

Many theorem provers have a ``small trusted kernel.'' (The term ``trusted'' here is possibly a misnomer, as it is not necessarily trusted or even trustworthy, but rather correctness-critical. But it is standard usage and we will use it throughout the paper.) The idea is that all trusted code be isolated in a relatively small corner of the program, where it can be inspected for correctness.

But if the goal is end-to-end formal correctness, this approach reveals certain flaws. For example, if the code is written in standard C++, then if undefined behavior is executed anywhere in the program, the entire run could be compromised, including the kernel. This means that correctness of the kernel depends not only on the kernel but also on the correctness (or at least lack of undefined behavior) of all code in the codebase.

We will take a more extreme approach: keep the ``small trusted kernel'' in its own process, leveraging the process boundary enforced by the operating system. This also means that the verifier is the complete application, so we can reasonably analyze the binary image directly rather than viewing it as a module in a larger code base in which other parts of the code are untrusted. This alternative approach can still be made to work with full formal correctness, but it requires the language to be formalized and the compiler to be proven correct, and the language must not have any ``unsafe'' features, which limits the capabilities of the untrusted part.

Having an external kernel also frees up everything else in the application from full formal correctness. The user interface to a theorem prover need not be formally correct, even if it contains its own proof checker. Of course it is undesirable for there to be bugs in this code, but errors here are not critical because the external verifier can always pick up the slack. Even if the prover interface is bug-ridden, as long as the exported artifact did not exercise those bugs, and the external verifier checks it, the resulting proof is still correct. The only trusted user interface code is the display of the theorems to prove, see \autoref{sec:arch}.

Unfortunately, most operating systems have ways to circumvent the process boundary, but there isn't much the child process can do to protect against this, so we have to add this as an assumption to the verifier run. For the most part, operating systems such as Linux protect processes from each other via virtual memory protection as long as the parent process (the shell or OS) does not give itself e.g. debugging permissions to the child process. A bare metal verifier can do better in this regard as there is no concurrent process to defend against (but there is firmware to contend with even then).

\subsection{The Metamath Zero architecture}\label{sec:arch}

\begin{figure}
\begin{center}
\begin{tikzpicture}[auto]
  \node at (-2, 0) [draw, fill=white, rectangle] (mm1) {\emph{foo}\texttt{.mm1}};
  \node at (.8, 0) [coordinate] (pt) {};
  \node at (2, 0.5) [draw, fill=white, rectangle] (mm0) {\uline{\emph{foo}\texttt{.mm0}}};
  \node at (2, -0.5) [draw, fill=white, rectangle] (mmb) {\emph{foo}\texttt{.mmb}};
  \node at (4, 0) [fill=white] (verify) {\uline{verifier}};
  \node at (4, 1) [fill=white] (display) {\uline{display}};
  \draw (mm1) -- node {compiler}(pt);
  \draw[->] (pt) |- (mm0);
  \draw[->] (pt) |- (mmb);
  \draw[->] (mm1) edge [loop above] node {editor} ();
  \draw[->] (mm0) -| (display);
  \draw[->] (mm0) -| (verify);
  \draw[->] (mmb) -| (verify);

\end{tikzpicture}
\end{center}
\caption{The MM0 workflow. \uline{Underlined} components are trusted.}
\label{fig:arch}
\end{figure}
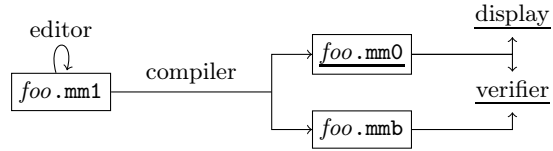
At this point, we have what we need to explain the overall architecture, depicted in \autoref{fig:arch}. The user writes proofs and programs in the proof assistant \textsf{MM1} (discussed in \autoref{sec:mm1}), receiving feedback on their proof. This live feedback is implemented in \textsf{MM1} and is not necessarily reliable (which is to say, \textsf{MM1} is not meant to be formally verified), but the quick feedback loop helps with producing the proof. Once it is done, the \textsf{MM1} compiler produces an \texttt{.mmb} (proof) file, and either produces an \texttt{.mm0} (specification) file or checks the result against a given \texttt{.mm0} file. The \texttt{.mm0} file is a human readable file containing the statements of axioms and target theorems, while the \texttt{.mmb} file is a binary artifact containing the proof. The trusted \mmO\ verifier then reads the \texttt{.mm0} and \texttt{.mmb} files, and reports success if the \texttt{.mmb} file is a proof of the \texttt{.mm0} statements.

The trusted components in this architecture are the verifier, and the \texttt{.mm0} file containing the statements of the theorems. Additionally one has to trust that the text file is faithfully shown to the reader, and the reader understands the content of the file. As such, the \texttt{.mm0} file format balances human-readability with simplicity of the verifier implementation required to parse it and validate it against the data in the proof file.

Crucially, the \texttt{.mmb} file, which contains the actual proof, is not trusted, since it is the verifier's job to make sure that this proof is correct. In particular, we are interested to ensure that no possible \texttt{.mmb} file input can cause the verifier to misbehave, overwrite its own memory via an exploit, or accidentally break its own proof guarantees. In this way, the \texttt{.mmb} input is untrusted in the strongest sense, such as we might treat data from a possibly malicious agent on the internet.

The remainder of the paper discusses the various components of this process. \autoref{sec:logic} describes the logical framework in which theorems are proved, \autoref{sec:mm0} describes the specification format, \autoref{sec:mmb} describes the proof format, and \autoref{sec:mm1} discusses how \mmO\ proof objects can be generated. \autoref{sec:translation} shows work that has been done to connect \mmO\ to other proof languages. \autoref{sec:bootstrap} discusses progress towards proving the correctness theorem for the verifier implementation.

\section{The Metamath Zero logic}\label{sec:logic}
\subsection{Metamath}
As its name suggests, Metamath Zero is based on Metamath \cite{Meg19}, a formal system developed by Norman Megill in 1990. Its largest database, \texttt{set.mm}, is the home of over 23000 proofs in ZFC set theory. In the space of theorem prover languages, it is one of the simplest, by design.

The name ``Metamath'' comes from ``metavariable mathematics,'' because the core concept is the pervasive use of metavariables over an object logic. An example theorem statement in Metamath is
$$\vdash (\forall x\,(\varphi\to \psi)\to(\forall x\,\varphi\to \forall x\,\psi))$$
which has three ``free metavariables:'' $x$, $\varphi$, and $\psi$. $\varphi$ and $\psi$ range over formulas of the object logic (let us say first order logic formulas like $\forall v_0\,v_0=v_1$), and $x$ ranges over variables of the object logic (that is, $x$ can be $v_0$, $v_1$, \dots).

However, this object logic never appears in actual usage. Rather, a theorem is proved with these metavariables, and then it is later applied with the metavariables (simultaneously) substituted for expressions that will contain more metavariables. For example one could apply the above theorem with the substitution $\{x\mapsto y,\ \varphi\mapsto\forall y\,\varphi,\ \psi\mapsto x=y\}$ to get:
$$\vdash (\forall y\,(\forall y\,\varphi\to x=y)\to(\forall y\,\forall y\,\varphi\to \forall y\,x=y))$$
which again contains metavariables (in this case $x,y,\varphi$) that can be further substituted later.

One consequence of the fact that variables like $x$ are themselves ``variables ranging over variables'' is that in a statement like $\forall x\,x=y$, the variable $y$ may or may not be bound by the $\forall x$ quantifier, because $x$ and $y$ may be the same variable. In order to express that two variables are different, the language includes ``disjoint variable provisos'' $A\fresh B$, which may be used as preconditions in theorems and assert that variables $A$ and $B$ may not be substituted for expressions containing a common variable. This is usually seen in the special cases $x\fresh y$, asserting that $x$ and $y$ are not the same variable, and $x\fresh \varphi$, asserting that the substitution to $\varphi$ does not contain the variable that $x$ is substituted to.

When a theorem is applied, a substitution $\sigma$ of all the variables is provided, and for each pair of variables $A\fresh B$, it is checked that for every pair of variables $v\in\sigma(A), w\in \sigma(B)$, the disjoint variable condition $v\fresh w$ is in the context. (This is why the relation is termed a ``disjoint variable condition'': if $A\fresh B$ then the set of variables in substitutions to $A$ and $B$ must be disjoint.)

This is essentially the whole algorithm. There is no built in notion of free and bound variable, proper substitution, or alpha renaming --- these can all be defined in the logic itself. It turns out that this is not only straightforward to implement (which explains why there are 17 known Metamath verifiers written in almost as many languages), but the fundamental operation, substitution, is effectively string interpolation in the sense of \texttt{printf}, which can be done very efficiently on modern computers. As a result, Metamath boasts some of the fastest checking times of any theorem prover library; the reference implementation, \texttt{metamath.exe}, can check the \texttt{set.mm} database of ZFC mathematics in about 8 seconds, and the fastest checker, \texttt{smm}, has performed the same feat in 0.7 seconds (on a 2-core Intel i5 1.6GHz). (This is a reported number from an older machine on an older and smaller \texttt{set.mm}. We reran \texttt{smm} single threaded on a Intel i7 3.9 GHz and the latest version of \texttt{set.mm}, and obtained $927\pm28$ ms.)

\subsection{Shortcomings of Metamath}
The primary differences between Metamath and Metamath Zero lie in the handling of first order variables (``variables over variables'' from the previous section), expression parsing, and definitions, so some attention is merited to the way these are handled. In each case, Metamath chooses the simplest course of action, possibly at the cost of not making a statement as strong as one would like.

\subsubsection{Bundling}
As has been mentioned already, variables can alias, which leads to a phenomenon known as ``bundling'' in which a theorem might mean many different things depending on how the variables are substituted. For example, $\vdash\exists x\,x=y$ is an axiom in \texttt{set.mm} with no disjointness assumptions on $x$ and $y$. There are essentially two different kinds of object language assertions encoded here. If $i\ne j$, then $\vdash\exists v_i\,v_i=v_j$ asserts that there exists an element equal to $v_j$, and when the indices are the same, $\vdash\exists v_i\,v_i=v_i$ asserts that there exists an element that is equal to itself. As it happens, in FOL both of these statements are true, so we are comfortable asserting this axiom.

Nevertheless, there is no easy way to render this as a single theorem of FOL, except by taking the conjunction of the two statements, and this generalizes to more variables -- a bundled theorem containing $n$ variables with no disjointness condition is equivalent (in the intended semantics) to $B_n$ shadow copies of that theorem in FOL, where $B_n$ is the $n$th Bell number, counting the number of ways that $n$ elements can be partitioned into groups, depending on whether they are mapped to the same or different variables. The Bell numbers grow exponentially, $B_n=e^{O(n\ln n)}$, so this is at least a theoretical problem.

From the point of view of the Metamath user, this is not actually a problem -- this says that Metamath in theory achieves \emph{exponential compression} over more traditional variable handling methods, in which variables with different names are always distinct. However, it is a barrier to translations out of Metamath, because of the resulting exponential explosion.

However, this is not a problem in practice, because the theoretically predicted intricately bundled theorems aren't written. Usually all or almost all first order variables will be distinct from each other, in which case there is exactly one corresponding FOL theorem (up to alpha renaming). In order to ease translations, \mmO\ requires that all first order variables be distinct, and shoulders the burden of unbundling in the automatic $\mathsf{MM}\to\mathsf{MM0}$ translation (see \autoref{sec:translation}).

\subsubsection{Strings vs trees}
Metamath uses strings of tokens in order to represent expressions. That is, the theorem $\vdash(\varphi\to\varphi)$ is talking about the provability of the expression consisting of five tokens $[\texttt{(}, \texttt{ph}, \texttt{->}, \texttt{ph}, \texttt{)}]$, with the initial constant \texttt{|-} distinguishing this judgment from other judgments (for example $\vdash\varphi$ asserts that $\varphi$ is provable, while $\wff\varphi$ asserts that $\varphi$ is a well formed formula). The upshot of this is that parsing is trivial; spaces between tokens are mandatory so it is often as simple as \texttt{tokens = mm\_file.split(" ")}. This makes correctness of the verifier simpler because the Metamath specification lines up closely with the internal data representation.

However, this leads to a problem when interpreting expressions as formulas of FOL. The axioms that define the $\wff\varphi$ judgment can be interpreted as clauses of a context-free grammar, and when that grammar is unambiguous there is a one-to-one relationship between strings and their parse trees, which are identified with the proofs of $\wff\varphi$ judgments \cite{Carn16}. So in effect, parsing is not required because the parses are provided with the proof. But unambiguity of a context-free grammar, though true for \texttt{set.mm} \cite{unambiguous}, is undecidable in general, yet is soundness critical --- if parentheses were omitted in the definition of $\wff\varphi\to\psi$ (that is, if the formation rule for wffs included the clause ``if the strings $u$ and $v$ are wffs then $u,\mbox{`$\to$'},v$ is a wff''), there would be two parses of $\bot\to\bot\to\bot$, and by conflating them it is not difficult to prove a contradiction.

Metamath Zero uses trees (or more accurately dags, directed acyclic graphs) to represent expressions, which has some other side benefits for the proof format (see \autoref{sec:mmb}). This on its own is enough to prevent ambiguity from leading to unsoundness. However, this means that an \mmO\ verifier requires a dynamic parser for its operation, which we will discuss in more detail in \autoref{sec:mm0}.

\subsubsection{Definitions}
In Metamath, a definition is no more or less than an axiom. Generally a new definition begins with an axiom defining a new syntax constructor, for example $\wff\exists! x\,\varphi$, and an axiom that uses the $\leftrightarrow$ symbol to relate this syntax constructor with its ``definition,'' for example $$y\fresh x,\ y\fresh \varphi\quad \vdash \exists! x\,\varphi\leftrightarrow\exists y\,\forall x\,(\varphi\leftrightarrow x=y).$$

Once again, the correctness of these definitional axioms is soundness critical but not checked by the verifier. Definitions such as the above definition of $\exists!$ are conservative and eliminable (this is a metatheorem that can be proved outside Metamath), and by convention almost all definitions in \texttt{set.mm} haactuallyve a syntactic form like this, that is, a new constructor $P(\bar x)$ is introduced together with an axiom $y_i\fresh x_j,\ y_i\fresh y_j\vdash P(\bar x)\leftrightarrow\varphi(\bar x,\bar y)$, where the additional variables $\bar y$ (disjoint from $\bar x$ and each other) are all bound in the FOL sense.

This convention is sufficiently precise that there is a tool that checks these criteria, but this goes beyond the official Metamath specification, and only one of the 17 verifiers (the \texttt{mmj2} verifier) supports this check. This effectively means that MM verification in practice extends beyond the narrow definition of MM verification laid out in the standard.

Metamath Zero bakes in a concept of definition, which necessitates a simple convertibility judgment. It also requires an identification of variables that are ``bound in the FOL sense,'' which means that it can no longer completely ignore the notion of free and bound variables, at least when checking definitions.

\subsection{The \mmO\ formal system}\label{sec:formal}

\begin{figure}
  \begin{align*}
    x,\varphi,s,f::={}&\langle\mbox{ident}\rangle \qquad\rlap{\mbox{names, metavariables, sort names, constructor names}}\\
    \Gamma::={}&\cdot\mid\Gamma,\ x:s\mid\Gamma,\ \varphi:s\,\overline{x}&&\mbox{contexts}\\
    e::={}&x\mid \varphi\mid f\,\overline{e}&&\mbox{expressions}\\
    A::={}&e,\quad \Delta::=\overline{A}&&\mbox{statements}\\
    \delta::={}&\!\!\!\!\begin{aligned}[t]
      &\sort\,s&&\mbox{sorts}\\
      \mid {}&\mathsf{term}\,f(\Gamma):s\,\overline{x}&&\mbox{constructors}\\
      \mid {}&\mathsf{def}\,f(\Gamma):s\,\overline{x}=\overline{y:s'}.\;e&&\mbox{definitions}\\
      \mid {}&\mathsf{axiom}\,(\Gamma;\Delta\vdash A)&&\mbox{axioms}\\
      \mid {}&\mathsf{thm}\,(\Gamma;\Delta\vdash A)&&\mbox{theorems}
    \end{aligned}&&\mbox{declarations}\\
    E::={}&\overline{\delta}&&\mbox{environment}\\
  \end{align*}\vspace{-18pt}
  $$\boxed{(E)\ \Gamma\ctx}\qquad\dfrac{}{\cdot\ctx}\qquad
  \dfrac{\sort\,s\in E\quad \Gamma\ctx}{\Gamma,x:s\ctx}\qquad
  \dfrac{\sort\,s\in E\quad \Gamma\ctx\quad \overline{x\in\Gamma}}{\Gamma,\varphi:s\,\overline{x}\ctx}$$\vspace{3pt}
  $$\boxed{(E)\ \Gamma\vdash e:s}\qquad\dfrac{(x:s)\in \Gamma}{\Gamma\vdash x:s}\qquad
  \dfrac{(\varphi:s\,\overline{x})\in \Gamma}{\Gamma\vdash \varphi:s}\qquad
  \dfrac{(f(\Gamma'):s\,\overline{x})\in E\quad\Gamma\vdash \overline{e}::\Gamma'}{\Gamma\vdash f\,\overline{e}:s}$$\vspace{3pt}
  $$\boxed{(E)\ \Gamma\vdash \overline{e}::\Gamma'}\qquad\dfrac{}{\Gamma\vdash\cdot::\cdot}\hspace{15pt}
  \dfrac{\Gamma\vdash \overline{e}::\Gamma'\quad (y:s)\in\Gamma}{\Gamma\vdash (\overline{e},y)::(\Gamma',x:s)}\hspace{15pt}
  \dfrac{\Gamma\vdash \overline{e}::\Gamma'\quad \Gamma\vdash e':s}{\Gamma\vdash (\overline{e},e')::(\Gamma',\varphi:s\,\overline{x})}$$\vspace{3pt}
  $$\boxed{(E)\ \delta\ok}\qquad\dfrac{}{\sort\,s\ok}\qquad
  \dfrac{\sort\,s\in E\quad \Gamma\ctx\quad \overline{x\in\Gamma}}{\mathsf{term}\,f(\Gamma):s\,\overline{x}\ok}$$\vspace{1pt}
  $$\dfrac{\sort\,s\in E\hspace{9pt} \Gamma,\overline{y:s'}\ctx\hspace{9pt} \overline{x\in\Gamma}\hspace{9pt} \Gamma,\overline{y:s'}\vdash e:s\hspace{9pt} \mathrm{FV}_{\Gamma,\overline{y:s'}}(e)\subseteq\overline{x}}{\mathsf{def}\,f(\Gamma): s\,\overline{x}=\overline{y:s'}.\;e\ok}$$\vspace{1pt}
  $$\dfrac{\Gamma\ctx\quad \overline{\Gamma\vdash A:s}\quad \Gamma\vdash B:s'}{\mathsf{axiom}\,(\Gamma;\overline{A}\vdash B)\ok}\quad
  \dfrac{\mathsf{axiom}\,(\Gamma;\overline{A}\vdash B)\ok\quad \Gamma,\overline{y:s'};\overline{A}\vdash B}{\mathsf{thm}\,(\Gamma;\overline{A}\vdash B)\ok}$$\vspace{3pt}
  $$\boxed{E\env}\qquad\dfrac{}{\cdot\env}\qquad
  \dfrac{E\env\quad E\vdash\delta\ok}{E,\delta\env}$$
  \begin{align*}
    \mathrm{V}(x)&=\{x\}&
    \mathrm{FV}(x)&=\{x\}\\
    \mathrm{V}_\Gamma(\varphi)&=\overline{x}&
    \mathrm{FV}_\Gamma(\varphi)&=\overline{x}&&\mbox{where $(\varphi:s\,\overline{x})\in\Gamma$}\\
    \mathrm{V}(f\,\overline{e})&=\textstyle{\bigcup_i\mathrm{V}(e_i)}&
    \mathrm{FV}(f\,\overline{e})&=\underline{\mathrm{FV}}(\overline{e}::\Gamma')\cup \{e_i\mid \Gamma'_i\in \overline{x}\}&&\mbox{where $f(\Gamma'):s\,\overline{x}$}
  \end{align*}
  \begin{align*}
    \underline{\mathrm{FV}}(\cdot::\cdot)&=\emptyset\\
    \underline{\mathrm{FV}}((\overline{e},y)::(\Gamma',x:s))&=\underline{\mathrm{FV}}(\overline{e}::\Gamma')\\
    \underline{\mathrm{FV}}((\overline{e},e')::(\Gamma',\varphi:s\,\overline{x}))&=\underline{\mathrm{FV}}(\overline{e}::\Gamma')\cup(\mathrm{FV}(e')\setminus\{e_i\mid \Gamma'_i\in \overline{x}\})
  \end{align*}
  \caption{MM0 syntax and well formedness judgments. $\overline{\cdot}$ denotes iteration or lists, and $e_i$ denotes the $i$th element of $\overline{e}$. The $\Gamma\ctx$, $\Gamma\vdash e:s$, $\Gamma\vdash \overline{e}::\Gamma'$, and $\delta\ok$ judgments are parameterized over a fixed global environment $E$. $(f(\Gamma'):s\,\overline{x})\in E$ means there is a $\mathsf{term}$ or $\mathsf{def}$ in $E$ with this signature. See \autoref{fig:proof} for the definition of $\Gamma;\overline{A}\vdash B$.}
  \label{fig:rules}
\end{figure}

\begin{figure}
  $$\boxed{(E;\Gamma;\Delta)\;\vdash A}\qquad \dfrac{A\in\Delta}{\vdash A}\qquad
  \dfrac{\vdash A\equiv B\quad \vdash A}{\vdash B}$$\vspace{1pt}
  $$\dfrac{\begin{gathered}
    (\Gamma';\overline{A}\vdash B)\in E\qquad \Gamma\vdash\overline{e}::\Gamma'\qquad \forall i,\ \vdash A_i[\Gamma'\mapsto \overline{e}]\\
    \forall i\,j\,x,\ \Gamma_i=x\notin\mathrm{V}_{\Gamma'}(\Gamma_j)\to e_i\notin \mathrm{V}_\Gamma(e_j)
  \end{gathered}}{\vdash B[\Gamma'\mapsto \overline{e}]}$$\vspace{3pt}
  $$\boxed{(E;\Gamma;\Delta)\;\vdash e\equiv e'}\qquad \dfrac{}{\vdash e\equiv e}\qquad
  \dfrac{\vdash e\equiv e'}{\vdash e'\equiv e}\qquad
  \dfrac{\vdash e_1\equiv e_2\quad \vdash e_2\equiv e_3}{\vdash e_1\equiv e_3}$$\vspace{1pt}
  $$\dfrac{\forall i,\ {\vdash e_i\equiv e'_i}}{\vdash f\;\overline{e}\equiv f\;\overline{e'}}\qquad
  \dfrac{\begin{gathered}
    (\mathsf{def}\,f(\Gamma'):s\,\overline{x}=\overline{y:s'}.\;e')\in E\qquad
    \Gamma\vdash(\overline{e},\overline{z})::(\Gamma',\overline{y:s'})\\
    \forall i\,j,\ z_i\notin \mathrm{V}_\Gamma(e_j)\qquad
    \forall i\,j,\ i\ne j\to z_i\ne z_j
  \end{gathered}}{\vdash f\,\overline{e}\equiv e'[\Gamma',\overline{y:s'}\mapsto\overline{e},\overline{z}]}$$\vspace{-5pt}
  \begin{align*}
    x[\Gamma'\mapsto \overline{e}]&=e_i&&\mbox{where $x=\Gamma'_i$}\\
    \varphi[\Gamma'\mapsto \overline{e}]&=e_i&&\mbox{where $\varphi=\Gamma'_i$}\\
    (f\,\overline{e'})[\Gamma'\mapsto \overline{e}]&=f\,\overline{e'[\Gamma'\mapsto \overline{e}]}
  \end{align*}\vspace{-10pt}
  \caption{MM0 proof and convertibility judgments $\Gamma;\Delta\vdash A$ and $\Gamma;\Delta\vdash e\equiv e'$. The arguments $E,\Gamma,\Delta$ are fixed and hidden. $(\Gamma';\overline{A}\vdash B)\in E$ means that an axiom or theorem with this signature is in $E$, that is, $\mathsf{axiom}\,(\Gamma';\overline{A}\vdash B)\in E$ or $\mathsf{thm}\,(\Gamma';\overline{A}\vdash B)\in E$.}
  \label{fig:proof}
\end{figure}

\mmO\ is intended to act as a schematic metatheory over multi-sorted first order logic. This means that it contains \emph{sorts}, two kinds of \emph{variables}, \emph{expressions} constructed from \emph{term constructors} and \emph{definitions}, and \emph{axioms} and \emph{theorems} using expressions for their hypotheses and conclusion. Theorems have \emph{proofs}, which involve applications of other theorems and axioms. The remaining sections will go into more detail on each of these points.

\subsubsection{Sorts}
An \mmO\ file declares a (finite) collection of sorts. Every expression has a unique sort, and an expression can only be substituted for a variable of the same sort. There are no type constructors or function types, so the type system is finite.
(Higher order functions are mimicked using open terms, see \autoref{sec:term}.)

\subsubsection{Variables}
\mmO\ distinguishes between two different kinds of variables. One may variously be called names, first order variables or bound/binding variables. These play the role of ``variable variables'' from Metamath, and will be denoted in this paper with letters $x,y,z,\dots.$ They are essentially names that may be bound by quantifiers internal to the logic. ``Substitution'' of names is $\alpha$-conversion; expressions cannot be substituted directly for names, although axioms may be used to implement this action indirectly. The other kind of variable may be called a (schematic) metavariable or second order variable, and these may \emph{not} be bound by quantifiers; they are always implicitly universally quantified and held fixed within a single theorem, but unlike names, they may be directly substituted for an expression. We use $\varphi,\psi,\chi,\dots$ to denote schematic metavariables.

In FOL, notations like $\varphi(\bar x)$ are often used to indicate that a metavariable is explicitly permitted to depend on the variables $\bar x$, and sometimes but not always additional ``parameter'' variables not under consideration. In \mmO, we use a binder $\varphi:s\,\overline{x}$, where $s$ is the sort and $\overline{x}$ are the \emph{dependencies} of $\varphi$, to indicate that $\varphi$ represents an open term that may reference the variables $\overline{x}$ declared in the context. This is opposite the Metamath convention which requires mentioning all pairs of variables that are \emph{not} dependent, but it is otherwise a merely cosmetic change. Such a variable may also be glossed as a pre-applied higher order variable; for example a variable of type $\varphi:\mathsf{wff}\;x$ can be interpreted as a predicate $P:U\to\mathsf{bool}$ where every occurrence of $\varphi$ in the statement is replaced with $P\;x$.

\subsubsection{Term constructors}\label{sec:term}
Term constructor declarations are represented in \autoref{fig:rules} by $\mathsf{term}\,f(\Gamma):s\,\overline{x}$. A term constructor is like a  function symbol in FOL, except that it can bind variables if names are used in the arguments. Examples of term constructors are $\mathsf{imp}(\_:\mathsf{wff},\_:\mathsf{wff}):\mathsf{wff}$, which defines implication as a binary operator on the sort $\mathsf{wff}$ (which can be shortened to $\mathsf{imp}:\mathsf{wff}\Rightarrow\mathsf{wff}\Rightarrow \mathsf{wff}$), and $\mathsf{all}\;(x:\mathsf{var},\varphi:\mathsf{wff}\,x):\mathsf{wff}$, which defines the forall binder --- $\mathsf{all}\;x\;\varphi$ represents the FOL expression $\forall x,\varphi(x)$.

Using the rules in \autoref{fig:rules}, we can calculate that $\mathrm{FV}(\mathsf{all}\,y\,\psi)=\mathrm{FV}(\psi)\setminus\{y\}$ (which accords with the usual FOL definition of the free variables of $\forall y,\psi(y)$), and $\mathrm{V}(\mathsf{all}\,y\,\psi)=\{y\}\cup\mathrm{V}(\psi)$.  It is easy to see that $\mathrm{FV}(e)\subseteq \mathrm{V}(e)$ generally; that is, every free variable in an expression $e$ is present in $e$. Metamath, and Metamath Zero, take the somewhat unorthodox approach of using $\mathrm{V}$ instead of $\mathrm{FV}$ in the definition of an admissible substitution (the side condition $\forall i\,j\,x,\ \Gamma_i=x\notin\mathrm{V}_{\Gamma'}(\Gamma_j)\to e_i\notin \mathrm{V}_\Gamma(e_j)$ in the theorem application rule in \autoref{fig:proof}, which says in words that if $\Gamma_j$ is a variable in the context that is not declared to depend on $x$, then the substitution for $\Gamma_j$ cannot contain the name that is being substituted for $x$), but this is sound because $\mathrm{FV}(e)\subseteq \mathrm{V}(e)$: if $e$ does not contain any occurrence of $x$ then it clearly does not contain a free occurrence of $x$. This is done because $\mathrm{V}$ is faster to compute than $\mathrm{FV}$, and they are equally expressive, assuming the axioms support alpha conversion, because any expression $e$ with $x\in \mathrm{V}(e)\setminus\mathrm{FV}(e)$ is alpha-equivalent to an expression $e'$ such that $x\notin\mathrm{V}(e')$.

\subsubsection{Definitions}
Definitions, denoted by $\mathsf{def}\,f(\Gamma):s\,\overline{x}=\overline{y:s'}.\;e$ in \autoref{fig:rules}, are similar to term constructors in that they are expression constructors, but terms are axiomatic while definitions are conservative, and can be unfolded by the convertibility judgment $\vdash e\equiv e'$. One should read the definition as asserting $\Gamma,\overline{y:s'}\vdash f\,\overline{\gamma}:=e$, where $\overline{\gamma}$ is the list of variables in $\Gamma$. The variables $\overline{y}$ are all required to be bound in $e$, but they are added to the context anyway because in \mmO\ the context must contain all variables in $\mathrm{V}(e)$, because it does not expand when traversing binders. The convertibility rule for definitions in \autoref{fig:proof} substitutes both the variables in $\Gamma$ as well as the variables $\overline{y}$, which provides limited support for alpha renaming.

\subsubsection{Axioms and theorems}
Provable assertions are simply expressions of designated sorts. A general axiom or theorem is really an inference rule $\Gamma;\Delta\vdash A$, where $\Delta$ is a list of hypotheses and $A$ is a conclusion, and $\Gamma$ contains the variable declarations used in $\Delta$ and $A$. For example, the \L ukasiewicz axioms for propositional logic in this notation are:
\begin{align*}
  \varphi\;\psi:\mathsf{wff}&;\cdot\vdash \varphi\to\psi\to\varphi\\
  \varphi\;\psi\,\chi:\mathsf{wff}&;\cdot\vdash (\varphi\to\psi\to\chi)\to(\varphi\to\psi)\to(\varphi\to\chi)\\
  \varphi\;\psi:\mathsf{wff}&;\cdot\vdash (\neg\varphi\to\neg\psi)\to(\psi\to\varphi)\\
  \varphi\;\psi:\mathsf{wff}&;\ \varphi\to\psi,\ \varphi\vdash \psi
\end{align*}
Things get more interesting with the FOL axioms:
\begin{align*}
  x:\mathsf{var},\varphi\;\psi:\mathsf{wff}\,x&;\cdot\vdash\forall x\,(\varphi\to\psi)\to(\forall x\,\varphi\to\forall x\,\psi)\\
  x:\mathsf{var},\varphi:\mathsf{wff}&;\cdot\vdash\varphi\to\forall x\,\varphi
\end{align*}

Notice that $\varphi$ has type $\mathsf{wff}\,x$ in the first theorem and $\mathsf{wff}$ in the second, even though $x$ appears in both statements. This indicates that in the first theorem $\varphi$ may be substituted with an open term such as $x<2$, while in the second theorem $\varphi$ must not contain an occurrence of $x$ (not even a bound occurrence).

One may rightly point out that this restriction seems unnecessary, particularly as we no longer have Metamath's excuse that the logic has no concept of bound variable. The reason for this choice is twofold. First, this enables compatibility with both Metamath (which would reject such FV-admissible substitutions) and FOL and HOL (which use individual variables rather than names with bundling). Second, it is faster. Theorem application is the hottest loop in the verifier, which has to go through possibly millions of them in a large development, and having a fast path is extremely helpful for this purpose; most theorems don't need alpha renaming or proper substitution, so those that do can afford a few extra theorem applications, possibly auto-generated by the proof authoring tool, in order to perform the renaming in the logic.

\subsubsection{Proofs and convertibility}
Metamath has only the first and third rules of \autoref{fig:proof}: the hypothesis rule, and the application of a theorem after (direct) admissible substitution. Metamath Zero adds the second rule, which consists only of definition unfolding and compatibility rules. Alpha renaming is not directly available, because it is a nonlocal operation, but it can be simulated through the making of definitions (as well as built using theorems, as we endorsed in the previous section).

The rule for $\mathsf{thm}\,(\Gamma;\overline{A}\vdash B)\ok$ allows additional dummy variables $\overline{y:s'}$ to be used in the proof, as long as they do not appear in the statement ($\overline{A}$ and $B$ must not mention $\overline{y}$). This in particular implies that all sorts are nonempty (but see the \textsf{free} modifier in \autoref{sec:modifier}).

\subsection[The \texttt{.mm0} specification format]{The \texttt{.mm0} specification format\footnote{
  \url{https://github.com/digama0/mm0/blob/master/mm0.md}
  %ANON\texttt{mm0/mm0.md}
  }}\label{sec:mm0}
The \texttt{.mm0} file is responsible for explaining to the reader what the statement of all relevant theorems is. It closely resembles the axiomatic description of \autoref{sec:formal}, but with a concrete syntax. The two FOL axioms above are rendered like so:

\begin{verbatim}
axiom all_mono {x: var} (P Q: wff x):
  $ A. x (P -> Q) -> A. x P -> A. x Q $;
axiom all_vacuous {x: var} (p: wff): $ p -> A. x p $;
\end{verbatim}
assuming the sorts \texttt{var}, \texttt{wff}, the terms \texttt{imp} and \texttt{all}, and notations \texttt{->} and \texttt{A.} for them have been previously declared.

\subsubsection{Sort modifiers}\label{sec:modifier}
Sorts have modifiers that limit what roles they can play. These are enforced by the verifier but not strictly necessary for expressivity.
\begin{itemize}
  \item Every statement is required to have a \textsf{provable} sort, so that one can assert that if $x:\mathsf{nat}$ then $\vdash x$ is nonsense and not permitted.
  \item The $\mathsf{free}$ modifier asserts that a sort cannot be used as a dummy variable, in which case the sort may possibly be empty.
  \item The $\mathsf{strict}$ modifier asserts that the sort cannot be used as a name. This is useful for metavariable-only sorts like $\mathsf{wff}$.
  \item The $\mathsf{pure}$ modifier asserts that the sort has no expression constructors (terms or defs). This is useful for name-only sorts like $\mathsf{var}$.
\end{itemize}

\subsubsection{No proofs}
As its name implies, the \texttt{.mm0} specification file is only about specifying axioms and theorems, so it does not contain any proofs. This is an unusual choice for a theorem prover, although some systems like Mizar and Isabelle support exporting an ``abstract'' of the development, with proofs omitted.

The reason for this comes back to our breakdown of the purpose of the different components of the architecture in \autoref{sec:arch}. The correctness theorem we are aiming for here is akin to security with respect to an extreme threat model. As an illustration, suppose you are trying to encourage formalization of some theorem of interest, let's say Fermat's Last Theorem (FLT), and you organize a competition. You write FLT as a \texttt{.mm0} file, and open it up for the world to submit proof attempts as corresponding \texttt{.mmb} files. \emph{Even in the face of bad faith proof attempts}, even if you are receiving gigabytes-long machine learned proofs, you want the assurance that if the verifier accepts it, then the theorem is proved from the axioms you defined. (It would also be nice to know that the verifier cannot misbehave in other ways, such as leaking your data to the internet, and in practice a proof that the verifier is correct will have to establish the lack of general misbehavior anyway, since most kinds of misbehavior such as buffer overflows can potentially be exploited to trick the verifier to accept bad proofs.)

This may seem obvious, but it is surprisingly common for proof systems to contain convenience features that lead to security flaws, or proofs of contradiction, that ``no reasonable person would use'', making the tacit assumption that the proof author is reasonable. In practice this means that the proof itself must be inspected to ensure that none of these exploits have been used, or else a social system must be in place so that we can trust the authors; but this defeats the point of formalizing in the first place.

Since an \texttt{.mm0} file is a \emph{formalization target} or \emph{problem statement}, it does not require or even accept proofs of its statements directly inline. Axioms and theorems look exactly the same except for the keyword used to introduce them.

\subsubsection{Abstract definitions}
We can do something similar with definitions. A definition requires a definiens in \autoref{fig:rules}, but we can instead write a definition with no definiens, so that it looks just like a term declaration. This allows us to assert \emph{the existence} of a term constructor which satisfies any theorems that follow, which gives us a kind of abstraction. Sometimes it is easier to write down characteristic equations for a function rather than an explicit definition, especially in the case of recursive functions.

If we view the entire \texttt{.mm0} file as a single theorem statement of the metalogic, then this construction corresponds to a second order (constructive) existential quantifier, complementing the second order universal quantifiers that are associated to theorems with free metavariables.

\subsubsection{Local theorems and definitions}
Once one is committed to not proving theorems in the specification file, most dependencies go away. Theorems never reference each other, and only reference terms and definitions involved in their statements. So if focus is given to one theorem, then almost everything else goes away, and even in extreme cases it becomes quite feasible to write down everything up to and including the axiomatic framework in a few thousand lines. In the above example of FLT, the specification file must define the natural numbers and exponentiation, but certainly not modular forms. These are properly the domain of the proof file. (The degree to which this statement is accurate depends to some extent on the theorem. FLT is something of an extreme case in that the statement requires much less technology than the proof. However, there is usually at least a 10 to 1 factor between definitions and proofs and often much more; \texttt{set.mm} for example contains 29 times as many theorems as definitions, and the theorems are on average 4.7 times longer than definitions, so removing all the theorems amounts to a 99.27\% reduction in formal content.)

But that means that the proof file must have license to introduce its own definitions and theorems, beyond the ones described in the specification file (but \emph{not} sorts, term constructors, or axioms). And this is exactly the piece that is missing in Metamath: Forbidding new axioms is necessary in order to prevent a malicious proof author from assuming false things, but in MM that also means no new definitions, and that is an untenable expressivity limitation.

\subsubsection{Notation}
In the abstract characterization, we did not concern ourselves with notation, presuming that terms were constructed inductively as trees, but early testing of the concrete syntax revealed that no one likes to read piles of s-expressions, and readability was significantly impacted. The notation system was crafted so as to make parsing as simple as possible to implement, while still ensuring unambiguity, and allowing some simple infix and bracketing notations. Notations are enclosed in \texttt{\$} sentinels (as in \LaTeX) so that parsing can be separated into a static part (containing the top level syntax of the language) and a dynamic part (containing user notations for mathematical operations that have been defined).

The dynamic parser is a precedence parser, with a numeric hierarchy of precedence levels $0,1,2,\dots$ with an additional level $\mathsf{max}$, forming the order $\mathbb{N}\cup\{\infty\}$. ($\mathsf{max}$ is the precedence of atoms and parenthesized expressions.) Infix constants are declared with a precedence, and left/right associativity. (An earlier version of \mmO\ used nonassociative operators and a partial order for precedence levels, but this complicated the parser for no added expressivity. We recognize that overuse of precedence ordering can lead to miscommunication, but this is in the trusted specification file anyway, so the drafter must take care to be clear and use parentheses responsibly.)

General notations are also permitted; these have an arbitrary sequence of constants and variables, and can be used to make composite notations like \verb|sum_ i < n ai| as an approximation of $\sum_{i<n}a_i$. The only restriction on general notations to make them unambiguous is that they must begin with a unique constant, in this case \verb|sum_|. This is restrictive, but usually one can get away with a subscript or similar disambiguating mark without significantly hampering readability. (This may be relaxed in higher level languages, but recall that we are still in the base of the bootstrap here, so every bit of simplicity matters.)

Coercions are functions from one sort to another that have no notation. For example, if we have a sort of set expressions and another sort of class expressions, we might register a coercion $\mathsf{set}\to\mathsf{class}$ so that $x\in y$ makes sense even if $x$ and $y$ are sets and $x\in A$ is a relation between a set and a class. For unambiguity, the verifier requires that the coercion graph have at most one path from any sort to any other.

\section[The \texttt{.mmb} binary proof file]{The \texttt{.mmb} binary proof file\footnote{
  \url{https://github.com/digama0/mm0/tree/master/mm0-c}
  %ANON\texttt{mm0/mm0-c}
  }}\label{sec:mmb}
Having a precise language for specifying formal statements is nice, but it is most powerful when coupled with a method for proving those formal statements. We have indicated several times now design decisions that were made for efficiency reasons. How does \mmO\ achieve these goals?

From \autoref{sec:arch}, we can see that the only constraint on the \texttt{.mmb} format is that it somehow guides the verifier to validate that the input \texttt{.mm0} specification is provable. A useful model to keep in mind is that of a powerful but untrustworthy oracle providing hints whenever the verifier needs one, or a nondeterministic Turing machine that receives its nondeterminism from external input.

There are two fundamental principles that guide the design: ``avoid search,'' and ``don't repeat yourself.'' By spoon-feeding the verifier a very explicit proof, we end up doing a lot less computation, and by pervasively deduplicating, we can avoid all the exponential blowups that happen in unification. Using these techniques, we managed to translate \texttt{set.mm} into \mmO\ (see \autoref{sec:translation}) and verify the resulting binary proof file in $195\pm5$ ms (Intel i7 3.9 GHz, single threaded). (This is not a fair comparison in that we are not checking \texttt{set.mm} as is, we are adding a bunch of information and rearranging it to be faster to check, and observing that the result is faster to check. But in a sense that's the point.) While \texttt{set.mm} is formidable, at 34 MB / 590 kLOC, we are planning to scale up to larger or less optimized formal libraries to see if it is competitive even on more adversarial inputs.

\subsection{High level structure}
The proof file is designed to be manipulated in situ; it does not need to be processed into memory structures, as it is already organized like one. It contains a header that declares the sorts, and the number of terms/defs and axioms/theorems, and then links to the beginning of the term table and the theorem table, and the declaration list.

The term table and theorem table contain the statements of all theorems and the types of all term constructors. These tables are consulted during typechecking, and the verifier uses a counter as a sliding window into the table to mark what part of the table has been verified (and thus is usable). This means that a term lookup is generally a single indexed memory access, usually in cache, which makes type checking for expressions ($\Gamma\vdash e:s$) extremely fast in practice.

Variable names, term names, and theorem names are all replaced as identifiers with indexes into the relevant arrays. All strings are stored in an index that is placed at the end of the file, linked to from the header, and not touched by the verifier except when it wants to report an error. It is analogous to debugging data stored in executables --- it can be stripped without affecting anything except the quality of error reporting.

A term entry contains a table of variable declarations (the context $\Gamma$ and the target type $s\,\overline{x}$) followed by a unify stream for definitions, and a theorem entry contains a table of variable declarations (the context $\Gamma$), followed by a unify stream (\autoref{sec:stream}).

\begin{figure}
  \begin{align*}
    \sigma ::={}&e\mid {{\vdash} A}\mid e\equiv e'\mid e\coconv e'&&\mbox{stack element}\\
    H,S,U,K::={}&\overline{\sigma}&&\mbox{heap, stack, unify heap, unify stack}\\
    \Delta::={}&\overline{A}&&\mbox{hypothesis list}
  \end{align*}
  \begin{align*}
    &\mathsf{Save}{:}&H;S,\sigma&\hto H,\sigma;S,\sigma\\[-7pt]
    &\mathsf{Term}\;f{:}&S,\overline{e}&\hto S,e'&&\!\!\!\!\left(\begin{gathered}
      f:\Gamma'\Rightarrow s\,\overline{x},\quad \vdash\overline{e}::\Gamma'\\[-3pt]
      e':=\alloc(f\,\overline{e}:s)
    \end{gathered}\right)\\[-4pt]
    &\mathsf{Ref}\;i{:}&S&\hto S,e&&(e:=H[i])\\
    &\rlap{$\mathsf{Dummy}\;s{:}$}&H;S&\hto H,e;S,e&&(e:=\alloc(x:s),\  x\mbox{ fresh})\\
    &\mathsf{Thm}\;T{:}&S,\bar{e}^*,A&\hto S',{{\vdash}A}&&(\mathrm{Unify}(T){:{}}S;\overline{e};A\uto S')\\
    &\mathsf{Hyp}{:}&\Delta;H;S,A&\hto \Delta,A;H,{{\vdash}A};S\\
    &\mathsf{Conv}{:}&S,A,{{\vdash}B}&\hto S,{{\vdash}A},\rlap{$A\coconv B$}\\
    &\mathsf{Refl}{:}&S,e\coconv e'&\hto S&&(e=e')\\
    &\mathsf{Symm}{:}&S,e\coconv e'&\hto S,e'\coconv e\\
    &\mathsf{Cong}{:}&S,f\,\overline{e}\coconv f\,\overline{e'}&\hto S,\overline{e\coconv e'}^*\\
    &\mathsf{Unfold}{:}&S,f\,\overline{e},e'&\hto S',e'\coconv e''&&
      (\mathrm{Unify}(f){:{}}S;\overline{e};e'\uto S',f\,\overline{e}\coconv e'')\\
    &\rlap{$\mathsf{ConvCut}{:}$}&S,e,e'&\hto S,\rlap{$e\equiv e',e\coconv e'$}\\
    &\rlap{$\mathsf{ConvRef}\;i{:}$}&S,e\coconv e'&\hto S&&(H[i]=e\equiv e')\\
    &\rlap{$\mathsf{ConvSave}{:}$}&H;S,e\equiv e'&\hto H,e\equiv e';S\\[7pt]
    &\mathsf{USave}{:}&U;K,\sigma&\uto U,\sigma;K,\sigma\\
    &\rlap{$\mathsf{UTerm}\;f{:}$}&K,f\,\overline{e}&\uto K,\overline{e}\\
    &\mathsf{URef}\;i{:}&U;K,e&\uto U;K&&(U[i]=e)\\
    &\rlap{$\mathsf{UDummy}\;s{:}$}&U;K,x&\uto U,x;K&&(x:s)\\
    &\mathsf{UHyp}{:}&S,{\vdash A};K&\uto S;K,A
  \end{align*}
  \caption{Proof stream and unify stream opcodes and their operational semantics. Proof steps have the form $C:\Delta;H;S\hto\Delta';H';S'$, and unify steps have the form $C:S;U;K\hto S';U';K'$, but values that do not change are suppressed. $\overline{e}^*$ denotes the reverse of $\overline{e}$.}
  \label{fig:stream}
\end{figure}

\subsection{The declaration list}\label{sec:stream}
After the term and theorem tables is the the declaration list, which validates each declaration in the \texttt{.mm0} file, possibly interspersed with additional definitions and theorems. This data is processed in a single pass, and contains in particular proofs of theorems. The global state of the verifier is very small; it need only keep track of how many terms, theorems, and sorts have been verified so far, treating some initial segment of the input tables as available for use and the rest as inaccessible. Because terms and theorems are numbered in the same order they appear in the file, when a theorem appears in the declaration list it is always the one just after the current end of the theorem table.

There are two kinds of opcode streams, proof streams and unify streams. Unify streams appear only in the term and theorem tables, and are used when a theorem is referenced or a definition is unfolded. Proof streams appear in the declaration list and provide proofs for theorems.

During a proof, the verifier state consists of a store (a write-once memory arena that is cleared after each proof) which builds up pointer data structures for constructed expressions, a heap $H$, and a stack $S$. A stack element $\sigma$ can be either an expression $e$ or a proof $\vdash A$, both of which are simply pointers into the store where the relevant expression is stored. The nodes themselves store the head and sort of the expression: $x:s$, $\varphi:s$, or $f\;\overline{e}:s$, as well as precalculating $\mathrm{V}(e)$ ($\mathrm{FV}(e)$ when constructing definition expressions). (There are also two kinds of convertibility proof that can be on the stack, discussed below.)

% Roughly speaking, a proof stream stores proofs in postfix order (RPN), so for example the proof $T_1(x,T_2(y),f(z))$ would be expressed as $\mathsf{Ref}\ x$, $\mathsf{Ref}\ y$, $\mathsf{Thm}\ T_2$, $\mathsf{Ref}\ z$, $\mathsf{Term}\ f$, $\mathsf{Thm}\ T_1$. A notable difference from Metamath is that a theorem with $n$ variables and $m$ hypotheses takes $m+n+1$ arguments off the stack, where the additional argument is the statement of the theorem to be proved. (Metamath only requires the $m+n$ arguments, and if first order unification is used only the $m$ subproofs are truly required to reconstruct the proof.)

% By contrast, the unify stream stores expressions (theorem hypotheses and conclusions, and definition bodies) in \emph{prefix} order; so for example $g(x,f(z))$ would be stored as $\mathsf{UTerm}\ g$, $\mathsf{URef}\ x$, $\mathsf{UTerm}\ f$, $\mathsf{URef}\ z$. The reason for this difference in convention is that the proof stream, which is responsible for constructing proofs of intermediate statements, works by \emph{building up} expressions, while the unify stream, which is responsible for proving facts of the form $e_1[\Gamma\mapsto \overline{e}]=e_2$ (for fixed $e_1,\Gamma$ and variable $\overline{e},e_2$), works by \emph{deconstructing} expressions, repeatedly matching the head of $e_1$ and pushing the pieces using $\mathsf{UTerm}$.

A declaration in the list is an opcode for the kind of declaration, a pointer to the next declaration (for fast scanning and parallelization), and some data depending on what kind of declaration it is:
\begin{itemize}
  \item Sorts and terms just mark the next item in the declaration table as valid for use.
  \item A definition $\mathsf{def}\;f(\Gamma):s\,\overline{x}=\overline{y:s'}.\;e$ reads a proof stream $\mathrm{Proof}(f){:{}}\cdot;\Gamma;\cdot\hto\cdot;\ \Gamma,\overline{y:s'};\ e$ (which is to say, it initializes the heap with $\Gamma$ and an empty stack, and expects a single expression $e$ on the stack after executing $\mathrm{Proof}(f)$), and then checks that $\mathrm{Unify}(f)$, the corresponding element of the declaration table, satisfies $\mathrm{Unify}(f){:{}}\cdot;\overline{y:s'}^*;e\uto \cdot;\Gamma';\cdot$.
  \item A theorem or axiom $T:(\Gamma,\Delta\vdash A)$ reads a proof stream $\mathrm{Proof}(T){:{}}\cdot;\Gamma;\cdot\hto \Delta^*;\ \Gamma,\overline{y:s'};\ {{\vdash} A}$ (for axioms, the stack at the end holds $A$ instead of ${\vdash} A$), and then checks $\mathrm{Unify}(T){:{}}\cdot;{{\vdash}\Delta};\ \Gamma;\ A\uto \cdot;\Gamma';\cdot$.
\end{itemize}

In short, we build up an expression using the $\mathrm{Proof}(f)$ proof stream, and then check it against the expression that is in the global space using $\mathrm{Unify}(f)$, so that we can safely reread it later.

At the beginning of a proof, the heap is initialized with expressions for all the variables. An opcode like $\mathsf{Term}\;f$ will pop $n$ elements $\overline{e}$ from the stack, and push $f\;\overline{e}$, while $\mathsf{Ref}\;i$ will push $H[i]$ to the stack. The verifier is arranged such that no expression is always accessed via backreference if it is constructed more than once, so equality testing is always $O(1)$.

The opcode $\mathsf{Thm}\;T$ pops $\overline{e}$ from the stack (the number of variables in the theorem), pops $B'$ from the stack (the substituted conclusion of the theorem), then calls a \emph{unifier} for $T$, stored in the theorem table for $T$, which is another sequence of opcodes. This will pop some number of additional $\overline{\vdash A'}$ assumptions from the stack, and then $\vdash B'$ is pushed on the stack.

The unifier is responsible for deconstructing $B'$ and proving that $B[\Gamma\mapsto \overline{e}]=B'$, where $B$ and $\Gamma$ are fixed from the definition of $T$, and $\overline{e}$ and $B'$ are provided by the theorem application. It has its own stack $K$ and heap $U$; the unify heap is the incoming substitution, and the unify stack is the list of unification obligations. For example $\mathsf{URef}\;i$ pops $e$ from the stack and checks that $U[i]=e$, while $\mathsf{UTerm}\;f$ pops an expression $e$ from the unify stack, checks that $e=f\;\overline{e'}$, and then pushes $\overline{e'}$ on the stack (in reverse order). The appropriate list of opcodes can be easily constructed for a given expression by reading the term in prefix order, with \textsf{UTerm} at each term constructor and \textsf{URef} for variables. The \textsf{UHyp} instruction pops $\vdash A'$ from the main stack $S$ and pushes $A'$ to the unify stack $K$; this is how the theorem signals that it needs a hypothesis.

Convertibility is handled slightly differently than in the abstract formalism. Most of the convertibility rules are inverted, working with a co-convertibility hypothetical $e\coconv e'$. In the absence of $e\coconv e'$ judgments on the stack, the meaning of the stack is that all $\vdash A$ statements in it are provable under the hypotheses $\Delta$, but $S,e\coconv e'$ means that if $e\equiv e'$ is provable, then the meaning of $S$ holds. So for instance, the $\mathsf{Conv}$ rule $S,A,{{\vdash} B}\ \hto\ S,{{\vdash} A},A\coconv B$ says that from $\vdash B$, we can deduce that if $\vdash A\equiv B$ is provable, then $\vdash A$ holds, which is indeed the conversion rule.

The reason for this inversion is that it makes most unfolding proofs much terser, since all the terms needed in the proof have already been constructed, and the $\mathsf{Refl}$ and $\mathsf{Cong}$ rules only need to deconstruct those terms.

The $\mathsf{ConvCut}$ rule is not strictly necessary, but is available in accordance with the ``don't repeat yourself'' principle. It allows for an unfolding proof to be stored and replayed multiple times, which might be useful if it is a frequently appearing subterm.

The handling of memory is interesting in that all allocations are ``controlled by the user'' in the sense that they happen only on $\mathsf{Term}\;f$ and $\mathsf{Dummy}\;s$ steps.  (Note that ``the user'' here is really the compiler, since the \texttt{.mmb} format is how the compiler communicates to the verifier.) Because proof streams are processed in one pass, that means that every allocation in the verifier can be identified with a particular opcode in the file.

But the biggest upshot of letting the user control allocation is that they have complete control over the result of pointer equality. That is, whenever a statement contains a subterm multiple times, for example $g(f(x),f(x))$, the user can arrange the proof such that these subterms are always pointers to the same element on the heap (in this example, $\mathsf{Ref}\;x$, $\mathsf{Term}\;f$, $\mathsf{Save}$, $ \mathsf{Ref}\;1$, $\mathsf{Term}\;g$, assuming that the $\mathsf{Save}$ puts $f(x)$ at index 1). This would not be possible without hash-consing if the verifier ``built expressions on its own volition'' in the course of performing substitution or applying theorems. As such, the verifier can simply \emph{require} that every term be constructed at most once (or at least, any expressions that participate in an equality test should be identified), and then expression equality testing in steps like $\mathsf{URef}$ and $\mathsf{Refl}$ is always constant time.

An earlier version of the verifier actually put the data that would otherwise need to be allocated into the instruction itself (i.e. the instruction might be $\mathsf{Term}(f,\overline{e},\overline{x})$, and the verifier is responsible for checking that $\mathrm{V}(f\;\overline{e})=\overline{x}$). However, this wastes a lot of space (the $\mathrm{V}(e)$ slots are typically 8 bytes) for ephemeral data. Putting too much data into the proof file means more IO to read it, which can cancel the performance benefits of not having to allocate memory. The memory highwater is under 1 megabyte even after reading the largest proofs in \texttt{set.mm} (which deliberately includes a few stress test theorems), so memory usage doesn't seem to be a major issue. Nevertheless, it is useful to note that by encoding the heap and stack in the instruction stream, it is possible to perform verification with $O(1)$ writable memory, streaming almost all of the proof.

Verification is not quite linear time, because each $\mathsf{Thm}\;T$ instruction causes the verifier to read $\mathrm{Unify}(T)$, which is approximately as large as the (deduplicated) statement of $T$. It is $O(mn)$ where $n$ is the length of the proof and $m$ is the length of the longest theorem statement, but the statements that exercise the quadratic worst case are rather contrived; they require \emph{one theorem statement} to be on the same order as the whole file. An example would be if we have a theorem $T:(a:\mathsf{nat}\vdash a\cdot \bar n=0)$, where $\bar n$ is a large unary numeral $S^n(0)$, and we have a proof that constructs and discards $T(0),\dots,T(\bar n)$, and then proves a triviality. This requires only $O(n)$ to state (because there are $O(n)$ expression subterms in the full proof), but each application of $T(i)$ requires $O(n)$ to verify because it must match the large term $\bar n$ in the statement of $T$, resulting in $O(n^2)$ overall. However, it should be emphasized that this is not a realistic workload; large theorem statements are very rare, and the large theorems that are used are rarely referenced multiple times in one proof, so for almost all reasonable proof libraries this achieves linear time verification.

\subsection{Compilation}

Fairly obviously, the \texttt{.mmb} format is not meant to be written by humans; instead it is ``compiled'' from source in some other human readable language. (The design is similar to, and indeed inspired by, high entropy machine code encodings.) The details of compilation depend on the form of this language, but the backend will probably be similar regardless. For the \textsf{MM1} compiler (see \autoref{sec:mm1}), after executing the high level tactics and programs, the result is an environment object in memory. Here expressions are stored in the usual functional way as trees (pointer data structures) with possible but not mandatory subexpression sharing, and proofs contain similar subproof sharing. For example, consider the following short \textsf{MM1} file and proof:
\begin{verbatim}
provable sort wff;
term im: wff > wff > wff; infixr im: $->$ prec 1;
axiom a1 (a b: wff): $ a -> b -> a $;
axiom mp (a b: wff): $ a -> b $ > $ a $ > $ b $;
theorem a1i (a b: wff) (h: $ a $): $ b -> a $ = '(mp a1 h);
\end{verbatim}
The proof \textsf{(mp a1 h)} is elaborated using first order unification to determine the necessary substitutions for term arguments, resulting in the elaborated proof \textsf{(mp a (im b a) (a1 a b) h)}. We also add an argument with the theorem statement of each intermediate step to obtain
$$\mbox{\textsf{(mp a (im b a) (a1 a b (im a (im b a))) h (im b a))}}.$$
(Already at this point we will have a lot of ``accidental'' subterm sharing, not shown in the printed s-expression, since it naturally appears as a result of first order unification and substitution.) We then perform hash-consing to ensure that every subterm has at most one index, also throwing in the expressions for the hypotheses by using a synthetic root node $\underline{\textsf{root}}$. We end up with a structure like so:
\begin{align*}
  \mathsf{let}\ 0:={}&\textsf{a},\ 1:=\textsf{b},\ 2:=\textsf{h},\ 3:=(\textsf{im}\ 1\ 0),\ 4:=(\textsf{im}\ 0\ 3),\\
  5:={}&(\textsf{a1}\ 0\ 1\ 4),\ 6:=(\textsf{mp}\ 0\ 3\ 5\ 2\ 3)\ \mathsf{in}\ (\underline{\textsf{root}}\ 0\ 6)
\end{align*}
We inline all references that appear at most once:
\begin{align*}
  &\mathsf{let}\ 0:=\textsf{a},\ 1:=\textsf{b},\ 2:=\textsf{h},\ 3:=(\textsf{im}\ 1\ 0)\\
  &\mathsf{in}\ (\underline{\textsf{root}}\ 0\ (\textsf{mp}\ 0\ 3\ (\textsf{a1}\ 0\ 1\ (\textsf{im}\ 0\ 3))\ 2\ 3))
\end{align*}
And now we can produce a proof stream by traversing this expression in postfix order, recursing into a numbered reference if it is the first appearance of the number. The numbers 0-1 in this case are already on the heap at the beginning of the stream because they are in the context.
\begin{align*}
  &\quad\mathsf{Ref}\;0,\mathsf{Hyp}^2,\\
  &\quad\quad\mathsf{Ref}\;0,\\
  &\quad\quad\mathsf{Ref}\;0,\mathsf{Ref}\;1,\mathsf{Term}\;\mathsf{im},\mathsf{Save}^3,\\
  &\quad\quad\mathsf{Ref}\;0,\mathsf{Ref}\;1,(\mathsf{Ref}\;0,\mathsf{Ref}\;3,\mathsf{Term}\;\mathsf{im}),\mathsf{Thm}\;\mathsf{a1},\\
  &\quad\quad\mathsf{Ref}\;2,\\
  &\quad\quad\mathsf{Ref}\;3,\\
  &\quad\mathsf{Thm}\;\mathsf{mp}.
\end{align*}
(The indentation and grouping is used here to indicate the tree structure, but the actual output is a plain list. The superscripts on \textsf{Hyp} and \textsf{Save} indicate what heap ID was associated to them. This is known by the compiler because the heap size goes up by one on each heap-modifying instruction and is initialized to $2$ at the start, because there are two variables $a,b$ in the theorem statement.)

The unify stream is similarly obtained by hash-consing the expression \textsf{(\underline{root} (im b a) a)} containing the statement of the theorem (note that the hypotheses come in reverse order, after the conclusion) and writing the result in prefix order:
$$\mathsf{UTerm}\;\mathsf{im},\ \mathsf{URef}\;1,\ \mathsf{URef}\;0,\ \mathsf{UHyp},\ \mathsf{URef}\;0.$$
The overall \texttt{.mmb} file is thus produced by serializing the header, then the term and definition statements, then the theorem statements, and finally the declaration list which contains all statements in the order they were declared, with proof streams for terms, defs, axioms and proofs. While certainly more work than verification, the cost is not significantly different from compilation from regular programming languages.

\section[The \texttt{.mm1} proof authoring file]{The \texttt{.mm1} proof authoring file\footnote{
  \url{https://github.com/digama0/mm0/blob/master/mm0-hs/mm1.md}
  %ANON\texttt{mm0/mm0-hs/mm1.md}
  }}\label{sec:mm1}
So far, we have talked about the \mmO\ verifier, that receives a very explicit proof from some untrusted source. But in some sense this is the easy problem, when compared with the problem of getting proofs in any kind of formally specified language in the first place. In order to make this pipeline useful, we need a way to produce formal proofs, and that means a front end to complement the \mmO\ back end.

There are two principal methods for producing \texttt{.mm0}/\texttt{.mmb} pairs: Translate them from another language, or write in a language that is specifically intended for compilation to \mmO. (Translations are discussed in \autoref{sec:translation}.)

The \textsf{MM1} language has a syntax which is mostly an extension of \textsf{MM0} which allows providing proofs of theorems. There are currently two \textsf{MM1} compilers, \href{https://github.com/digama0/mm0/blob/master/mm0-hs/}{\texttt{mm0-hs}} written in Haskell and \href{https://github.com/digama0/mm0/blob/master/mm0-rs/}{\texttt{mm0-rs}} written in Rust, both of which provide verification, parsing and translation for all the \mmO\ family languages (the three formats mentioned in this paper, plus some debugging formats), compilation of \textsf{MM1} files to \textsf{MMB}, and a server compliant with the Language Server Protocol to provide editing support (syntax highlighting, live diagnostics, go-to-definition, hovers, etc.) for Visual Studio Code, extensible to other editors in the future.

For the bootstrapping project, we used \mmO\ to specify Peano Arithmetic (PA), and within this axiomatic system we defined the x86 instruction set architecture \cite{Carn19} and the \mmO\ formal system as defined in \autoref{sec:mm0}, to obtain an end-to-end specification from input strings, through lexing, parsing, specification well-formedness, type checking, and proof checking, relating it to the operation of an ELF binary file.

Of these, only the PA framework\footnote{
  \rlap{\url{https://github.com/digama0/mm0/blob/master/examples/peano.mm1}}
  %ANON\texttt{mm0/examples/peano.mm1}
} has been proved thus far, and the remainder are only \texttt{.mm0} specifications. But this already includes about 1000 theorems defining:
\begin{itemize}
  \item propositional logic,
  \item natural deduction style,
  \item first order logic over $\mathsf{nat}$,
  \item a second-order sort $\mathsf{set}$ that ranges over subsets of $\mathsf{nat}$ (this is a conservative extension because $\mathsf{set}$ is a $\mathsf{strict}$ sort; one cannot quantify over sets so they are just syntax sugar over wffs with one free variable),
  \item a definite description operator $\mathsf{the}:\mathsf{set}\to\mathsf{nat}$ (also a conservative extension, allowing the definition of functions like exponentiation from functional relations).
  \item numerals and arithmetic,
  \item The Cantor pairing function,
  \item (signed) integers,
  \item GCD, Bezout's lemma, the chinese remainder theorem,
  \item Exponentiation, primitive recursion,
  \item The Ackermann bijection, finite set theory,
  \item Functions, lambda and application,
  \item Lists, recursion and functions on lists.
\end{itemize}

None of this is particularly difficult, but it does cover the majority of the set-up work for doing metamathematics in PA. This \textsf{MM1} formalization is in part based on parts of \texttt{set.mm}, particularly in propositional logic and FOL, but \texttt{set.mm} is based on ZFC, not PA, and there is not much overlap. But it is enough to get the sense of the scalability of the approach. After compilation, verification takes $2\pm0.05$ ms using \texttt{mm0-c}, which makes sense since it is only a small fraction of the size of \texttt{set.mm}. Compilation is also quite competitive, at $306\pm 4$ ms using \texttt{mm0-rs}. Although it can barely be more than guesswork at this point, we don't anticipate the x86 verification part to be more than 100 times larger than this project (and that's a generous estimate), except possibly the compiler execution itself, which can skip the \textsf{MM1} interface and produce $\texttt{.mmb}$ directly.

Here we see an important reason for speed: the faster the server can read and execute the file, the faster the response time to live features like diagnostics that the user is relying on for making progress through the proof. We initially intended to add save points in between theorems so that we don't have to process the entire file on each keypress, as well as  theorem level parallelism, but the round trip time for diagnostics stayed under half a second throughout the development of \texttt{peano.mm1}, so it never became a sufficiently pressing problem to be worth implementation.

The \textsf{MM1} language also contains a Turing-complete meta-programming language based on Scheme. It is intended for writing small ``tactics'' that construct proofs. Besides a few small quality-of-life improvements, we used it to implement a general algorithm for proving congruence lemmas (theorems of the form $A = B\to f(A)=f(B)$) for all new definitions.

While \textsf{MM1} has a long way to go to compete with heavyweights in the theorem proving world like Coq, Isabelle, or Lean, we believe this to be an effective demonstration that even a parsimonious language like Metamath or \mmO\ can be used as the backend to a theorem prover, and ``all'' that is necessary is a bit of UI support to add features like a type system, a tactic language, unification, and inference; the mark of a good underlying formal system is that it gets out of your way and lets you say what needs to be said -- this is what we mean by ``expressivity.''

\subsection{Does MM1 generate MM0 files?}
The \textsf{MM1} language directly supports features for being able to generate \texttt{.mm0} files. This is one of the reasons why it has a similar syntax; if one deletes the proofs and all \textsf{local theorem}s, and additional extensions, then the result is basically a valid \texttt{.mm0} file.

Alternatively, one could write a \texttt{.mm0} file first, change the extension to \texttt{.mm1}, then ``fill it out'' progressively with proofs until the original specification is proved. This is what we did for \texttt{peano.mm0}. The approach of generating an \texttt{.mm0} file is similar to the ``abstract'' functionality of Mizar and Isabelle alluded to earlier. But a moment's consideration of \autoref{fig:arch} reveals a weakness of this approach: \emph{foo}\texttt{.mm1} is not trusted, but it generates a file \emph{foo}\texttt{.mm0} that is trusted. How is this? It is difficult to trust a build artifact that is hidden away.

We found it helpful to maintain \emph{both} \texttt{peano.mm0} and \texttt{peano.mm1}, even though they share a lot of common text. When they are both tracked by version control, it makes any changes to the axioms or statements much more obvious, drawing attention to the important parts. The relationship is formally checked, so we need not fear them falling out of alignment. Additionally, it is much easier to make the \texttt{.mm0} file look good (clear, unambiguous, well formatted) if it is manually written; much more effort must be put into a formatting tool to get a similar effect.

\section{MM0 as an interchange format}\label{sec:translation}
\mmO\ is a \emph{logical framework} in the sense that it doesn't prescribe any particular axioms or semantics. This makes it well suited for translations to and from other systems. A downside of this approach is that while correctness with respect to the formal system is well defined, \emph{soundness} becomes unclear in the absence of a fixed foundation. Instead, one gets several soundness theorems depending on what axioms are chosen and what semantics is targeted. General \mmO\ has a soundness theorem as well, similar to the Metamath soundness theorem \cite{Carn16}, but these models are rather unstructured. (There is a simple multi-sorted SOL model for \mmO, but it fails completeness.)However, for the short term, proof translation can function as a substitute for a soundness proof, and indeed, a proof translation amounts to building a class model of the source system in the target system.

Eventually, we hope to use \mmO\ to prove correctness of other theorem provers, and vice versa, and proof translations play an important role in this. There is a $O(n^2)$ problem with having $n$ mutually supporting bootstraps, as there are $n^2$ proofs to be done. But the proof of $A\vdash \mbox{`$A$ is correct'}$ is closely related to the proof of $B\vdash \mbox{`$A$ is correct'}$; if we had a method for translating proofs in $A$ to proofs in $B$, we would obtain the result immediately. Moreover, proof translations compose, so it only requires a spider-web of proof connections before we can achieve such a critical mass. (Of course, this is only enough to get each prover to agree that $A$ is correct. With $n$ verifiers we would need $n$ correctness proofs and an $O(n)$ network of translations to get the full matrix of correctness results.)

Our work in this area is modest, but it has already been quite helpful. Several times now we have mentioned verification of \texttt{set.mm} in \mmO, but this is a gigantic library that we would not have a hope of creating without a huge investment of time and effort. Instead, we map \textsf{MM} statements to \mmO, and then we obtain tens of thousands of \mmO\ theorems in one fell swoop, a huge data set for testing that we could not have obtained otherwise.

\subsection{Translating MM to MM0}\label{sec:frommm}
The Haskell verifier \texttt{mm0-hs} contains a \texttt{from-mm} subcommand that will convert Metamath proofs to MM0. Because of the similarity of the logics, the transformation is mostly cosmetic; unbundling is the most significant logical change. Whenever Metamath proves a theorem of the form $\vdash T[x,y]$ with no $x\fresh y$ assumption, we must generate two theorems, $\vdash T[x,x]$ and $\vdash T[x,y]$ (which implicitly assumes $x\fresh y$ in \mmO). In many cases we can avoid this, for example if $x$ and $y$ are not bound by anything, as in $\vdash x=y\to y=z\to x=z$, we can just make them metavariables instead of names, but some theorems require this treatment, like $\vdash \forall x\,x=y\to \forall y\,y=x$.

For definitions, we currently do nothing (we leave them as axioms), but we plan to detect MM style definitional axioms and turn them into \mmO\ definitions.

\subsection{Translating MM0 to HOL systems}\label{sec:tohol}
The \texttt{to-hol} subcommand translates MM0 into a subset of HOL in a very natural way. A metavariable $\varphi:s\,\overline{x}$ becomes an $n$-ary variable $\varphi:s_1\to\dots \to s_n\to s$, where $x_i:s_i$, and all occurrences of $\varphi$ in statements are replaced by $\varphi\,\overline{x}$. All hypotheses and the conclusion, are universally closed over the names, and the entire implication from hypotheses to conclusion is universally quantified over the metavariables.

For example, the axiom of generalization is
$$x:\mathsf{var},\varphi:\mathsf{wff}\,x;\ \varphi\vdash \mathsf{all}\,x\,\varphi,$$
which becomes
$$\forall\varphi:\mathsf{var}\to\mathsf{wff},\ (\forall x:\mathsf{var}, \vdash \varphi\,x)\Rightarrow{{}\vdash{}} \mathsf{all}\;(\lambda x:\mathsf{var},\,\varphi\,x)$$
after translation.

The actual output of \texttt{mm0-hs to-hol} is a bespoke intermediate language (although it has a typechecker), which is used as a stepping-off point to OpenTheory and Lean. One of the nice side effects of this work was that Metamath theorems in \texttt{set.mm} finally became available to other theorem provers. We demonstrate the utility of this translation by proving Dirichlet's theorem in Lean\footnote{
  \url{https://github.com/digama0/mm0/blob/master/mm0-lean/mm0/set/post.lean}
  %ANON\texttt{mm0/mm0-lean/mm0/set/post.lean}
}, using the number theory library in Metamath for the bulk of the proof and post-processing the statement so that it is expressed in idiomatic Lean style. Very little of the Lean library was used for the proof. We only needed to show things like an isomorphism between Lean's $\mathbb{N}$ and Metamath's $\mathbb{N}$, which follows because both systems have proved the universal property of $\mathbb{N}$.

\section{Bootstrapping}\label{sec:bootstrap}
There are a few components that go into bootstrapping a theorem prover. In short, what we want to do is prove a theorem of the form $\vdash{}$`\texttt{mm0.exe} is a piece of machine code that when executed given input $E$, terminates with exit code $0$ only if $E\env$'. In order to even write this statement down, we need:
\begin{enumerate}
  \item The definition of $E\env$, that is, the formalization of \autoref{sec:formal}.
  \item The definition of \texttt{mm0.exe}, that is, a compiled executable artifact that can act as a verifier in the manner of \texttt{mm0-c}.
  \item The definition of executing a piece of machine code, which requires the formalization of the semantics of the target architecture, in this case x86-64 Linux. (It's not actually a \texttt{.exe} file, it is an ELF file.)
\end{enumerate}
We have done parts 1 and 3, and avoided the need for part 2 by the use of an abstract \textsf{def}. For part 1, in addition to just the formal system, we also need to formalize parsing of \texttt{.mm0} files, since the input $E$ is a string (a sequence of bytes), which must first be parsed into the abstract syntax used in \autoref{sec:formal}. This is all done in the file \href{https://github.com/digama0/mm0/blob/master/examples/mm0.mm0}{\texttt{mm0.mm0}}\footnote{
  \url{https://github.com/digama0/mm0/blob/master/examples/mm0.mm0}
  %ANON\texttt{mm0/examples/mm0.mm0}
  }, which as the name suggests is a definition of the syntax and well formedness judgments of \mmO\ as a \mmO\ file.

Part 3 is done in \href{https://github.com/digama0/mm0/blob/master/examples/x86.mm0}{\texttt{x86.mm0}}\footnote{
  \url{https://github.com/digama0/mm0/blob/master/examples/x86.mm0}
  %ANON\texttt{mm0/examples/x86.mm0}
}, which defines decoding of a subset of user-mode x86-64 instructions, the user-visible state of the architecture, and the execution semantics of these instructions. Although x86 is a notoriously large and complex instruction set architecture, a mere 25 or so instructions is sufficient to cover a program like \texttt{mm0-c}, which uses only integer registers and basic control flow.

One instruction is special: the \textsf{syscall} instruction allows performing IO operations, some of which go beyond the state of the x86 machine itself. To model this, we wrap the machine state with an IO state containing a stream of input waiting on \textsf{stdin}, and a stream of output so far on \textsf{stdout}. This allows us to view the entire program semantics as a (nondeterministic partial) function taking a list of bytes as input and producing a list of bytes as output, plus an exit code. The predicate \href{https://github.com/digama0/mm0/blob/775e6dd/examples/x86.mm0#L1230-L1236}{$\mathtt{succeeds}\ k\ i\ o$} in \href{https://github.com/digama0/mm0/blob/master/examples/x86.mm0}{\texttt{x86.mm0}}, for example, asserts that starting from initial state $k$ (in which the code has already been loaded), with input $i$, the machine will eventually reach some state $k'$, having consumed the entire input and produced output $o$, and return exit code $0$.

These parts are brought together into \href{https://github.com/digama0/mm0/blob/master/examples/x86_mm0.mm0}{\texttt{x86\_mm0.mm0}}\footnote{
  \url{https://github.com/digama0/mm0/blob/master/examples/x86_mm0.mm0}
  %ANON\texttt{mm0/examples/x86_mm0.mm0}
}, containing the short final theorem. We use \href{https://github.com/digama0/mm0/blob/master/examples/x86_mm0.mm0#L9}{\texttt{def Verifier: string;}} to declare an abstract definition of the verifier code itself (part 2 above), assert that it parses to a valid ELF file \texttt{VerifierElf}, and then the main theorems to prove are:
\begin{verbatim}
theorem Verifier_terminates (k s: nat):
  $ initialConfig VerifierElf k -> alwaysTerminates k s 0 $;
theorem Verifier_Valid (k s: nat):
  $ initialConfig VerifierElf k /\ succeeds k s 0 -> Valid s $;
\end{verbatim}
which assert that if the operating system loads \texttt{VerifierElf} into memory, resulting in initial state $k$, then on any input $s$, the program always terminates with no output (this is a bare verifier, which only produces success as an exit code), and if it succeeds, then $s$ is \texttt{Valid}, meaning that it parses as a \mmO\ file which is well formed, and all the theorems in the file are provable.

The proof file, and the formalization of \autoref{sec:mmb}, is nowhere to be found in this specification. While details about the proof format will certainly appear in the \texttt{Verifier} code (that is, \texttt{mm0-c}), and in the proof of these theorems, it is otherwise entirely hidden from the specification, because it is a ``secret'' input to the verifier. In fact, while the \mmO\ file is read from \textsf{stdin}, the \texttt{.mmb} file is provided as a command line argument and read using regular file IO. The way file IO is specified is that reading any file other than \textsf{stdin} produces a nondeterministic result: if it succeeds, the returned buffer can contain any sequence of bytes. Since the verifier is allowed to fail at any time, this is fine; the verifier simply checks that the proof file is a proof of the \mmO\ specification, and if so returns success.

Note that this specification is trivially satisfied by a verifier that always fails. It is not sufficient to merely have a verifier that is correct to the specification; one should also \emph{use} it on theorems of interest and \emph{observe} that the verifier reports success. In this case, we would want to use it on \texttt{x86\_mm0.mm0} to verify that it is correct. It is this interplay of the metalevel (the computer itself) and the logical model of the computer that is characteristic of bootstrapping systems.

\subsection{A proof producing compiler}
(We come now to future work. This is only a planned implementation.)

The previous section gives us a theorem statement, but how can we prove it? The general argument is that we have a piece of machine code that has been compiled from a higher level language such as C, and the code in this language has been carefully written to check that the proof file is valid. This is program verification, and it is a field all on its own, but we will use a few techniques to simplify the task.
\begin{itemize}
  \item The process of reducing assertions about a low level language to simpler assertions in a high level language is very similar to compilation. Many of the phases of a traditional compiler can be identified, for example allocating regions of the stack for local variables and register allocation. However in proof-producing compilation we have a second component to deal with, namely preservation of logical properties. This appears in traditional compilers as well, but usually only in the guise of behavior preservation. Lowering properties directly gives some added flexibility to the compiler.
  \item Rather than proving the compiler correct once and for all, the compiler can produce a piece of code and a proof that this code has the desired property in tandem. This is a strictly more general approach, and it allows us to perform various nondeterministic or untrusted operations in the middle of the proof. For example, algorithms for register allocation are notoriously difficult to prove correct, but they are easy to verify. By running such an untrusted algorithm and verifying the result, we obtain a shorter proof, with less effort, than if we were to prove the algorithm correct and then run the steps of the algorithm in a proof producing way.
  \item Another advantage of not proving the compiler correct once and for all is that we need not have a well-defined high level language at all. Defining the edge cases of a language can be one of the hardest parts, especially for a ``real'' language like C. Since the goal is only to compile a well behaved program, the compiler tactics can be fairly lax regarding edge cases, as long as the idioms in the target program are covered.
\end{itemize}

Because we are working inside a theorem prover, we have an ambient logic, and each intermediate language has to be represented in the logic. There are (at least) two possible ways to represent programs in the logic, which we shall call the ``syntactic'' and ``semantic'' approaches. In the syntactic approach, one defines a data type for program expressions, and a compilation function that manipulates these expressions. In the semantic approach, a program is a predicate on the set of all program behaviors, and program expressions are constructed from higher order functions in the logic. We use the semantic approach when possible.

This has a number of advantages. Because a program \emph{is} a property of behaviors, program correctness is built in; if we are able to compile the program then we have proven the theorem. It also allows us to use equality in the logic more usefully, since two programs with the same behavior are equal. Because the compiler is a tactic, a program at the meta-level, it is still capable of examining the structure of program expressions, but program expressions are not a closed class, which makes extensibility simpler; all that is necessary to define a new language construct is to add the \textsf{def} for it and teach the compiler how to compile the new construct, without changing the intermediate language.

One question is what the compiler's input language should look like. Since the compiler is a tactic, the input need not be expressed in the logic, but it must have a way to express computations and logical properties. Unfortunately, although compilers for languages such as C use logical reasoning extensively, this reasoning does not make it all the way to the surface syntax, and instead compilers try to infer properties using knowledge of undefined behaviors. While this has the major advantage that no annotation is necessary, it often leads to ``adversarial'' behavior when the program has a bug. Additionally, the kinds of properties that can be inferred in this way are quite limited.

Languages such as Dafny \cite{dafny} and Why3 \cite{why3} provide an input language with full first order logic, used inside the contracts of an imperative language. However, these rely on SMT solvers for proving verification conditions, which makes it difficult to connect them to a proof-producing back-end (most SMT solvers are not proof-producing). Generally code extraction in these tools is unverified, but because the front end has a well defined semantics, it is possible that future work can verify the back end to produce end to end correctness proofs without modifying user code. For the present work we will aim for a small subset of these languages.

\section{Related work}\label{sec:related}
The idea of a bootstrapping theorem prover is not new. There are a number of notable projects in this space, many of which have influenced the design of \mmO. However, none of these projects seem to have recognized (in words or actions) the value of parsimony, specifically as it relates to bootstrapping.

At its heart, a theorem prover that proves it is correct is a type of circular proof. While a proof of correctness can significantly amplify our confidence that we haven't missed any bugs, we must eventually turn to other methods to ground the argument, and direct inspection is always the fallback. But the effectiveness of direct inspection is inversely proportional to the size of the artifact, so the only way to make a bootstrap argument more airtight is to make it smaller.

The most closely related projects, in terms of bootstrapping a theorem prover down to machine code, are CakeML and Milawa.
\begin{itemize}
\item CakeML \cite{Kumar14} appears to be the most active bootstrapping system today. The bootstrap consists of two parts: CakeML is a compiler for ML that is written in the logic of HOL4 \cite{slind08}, and HOL4 is a theorem prover written in ML. Since the completion of the bootstrap in 2014, the CakeML team have expanded downward with \emph{verified stacks} \cite{loow19}, formalizing the hardware of an open source processor design they could implement using an FPGA.

Unfortunately, it seems that the bootstrap is not complete in the sense that the ML that CakeML supports is not sufficient for HOL4, and while a simpler kernel, called Candle, has been implemented in CakeML, it supports a variant of HOL Light, not HOL4, and cannot handle many of the idioms used in the correctness proof of the CakeML compiler. Furthermore, the compiler correctness proof takes on the order of 14 hours to run, and while we do not yet have reliable figures, we project that the \mmO\ toolchain will be able to beat this figure by 3--5 orders of magnitude (or 5--7 counting only the verification time of the completed proof).

\item Milawa \cite{davis09} is a theorem prover based on ACL2 developed for Jared Davis's PhD thesis, which starts with a simple inspectable verifier A which proves the correctness of a more powerful verifier B, which proves verifier C and so on. After another 12 steps or so the verifier becomes practical enough to be able to prove verifier A correct. This project was later extended by Magnus Myreen to \emph{Jitawa} \cite{myreen11}, a Lisp runtime that was verified in HOL4 and can run Milawa. Although this isn't exactly a bootstrap, it is an instance of bootstrap cooperation (to the extent that CakeML/HOL4 can be considered a bootstrap), of the sort we described in \autoref{sec:translation}.
\end{itemize}

There are a few other projects that have done bootstraps at the logic level. The original version of Milawa has this characteristic, since it does not go down to machine code but rather starts from a Lisp-like programming language with proof capabilities and uses this language to write a type checker for its own language. This means that issues such as compiling to an architecture at the back end, and verified parsing at the front end, don't come up and have to be trusted.
\begin{itemize}
  \item ``Coq in Coq'' by Bruno Barras (1996) \cite{barras1996coq} is a formalization of the Calculus of Constructions (CC) and a typechecker thereof in Coq, which can be extracted into an OCaml program. Here it is not Coq itself that is being verified but rather an independent kernel; moreover Coq implements not CC but CIC (the calculus of \emph{inductive} constructions), and of course many inductive types are used in the construction of the typechecker, so this fails to ``close the loop'' of the bootstrap.

  \item John Harrison's ``Towards self-verification of HOL Light'' (2006) \cite{harrison-holhol} writes down a translation of the HOL Light kernel (written in OCaml) in HOL Light, and proves the soundness of the axiom system with respect to a set theoretical model. This is the earliest example we know of a theorem prover verifying its own implementation, but it leaves off verification of OCaml (quite to the contrary, it is explicitly mentioned in the paper that it is possible to violate soundness using string mutability), and the translation from OCaml code to HOL Light definitions is unverified and slightly nontrivial in places.

  \item ``Coq Coq Correct!'' (2019) \cite{sozeaucoq} improves on ``Coq in Coq'' by verifying a typechecker for PCUIC (the polymorphic, cumulative calculus of inductive constructions), which is a much closer approximation to Coq. It still lacks certain features of the kernel such as the module system and some advanced kinds of inductive types, and some core components like the guard condition are left undefined by the specification. However, the implemented subset of Coq is at least expressive enough to contain the formalization itself. Sadly, the typechecker is not fast enough in practice to be able to typecheck its own formalization.
\end{itemize}

The \mmO\ project draws from ideas in a number of fields, most of which have long histories and many contributors.
\begin{itemize}
  \item Code extraction is the process of taking a definition in the logic and turning it into executable code, usually by transpilation to a traditional compiled language. Isabelle/HOL \cite{haftmann2013code} can target SML, Scala, OCaml, and Haskell; HOL4 can target OCaml, and Coq \cite{letouzey2008extraction} can target OCaml and Haskell. This is the most popular way of having simultaneously an object to prove properties about, and a program that is reasonably efficient. However, as argued in \cite{kumar2018software}, we believe that this leaves large gaps in the verified part, since the extraction function must be trusted as well as the target language's compiler.

  \item ISA specification is becoming more commonplace. \cite{k-x86} is a complete formal specification of the user level Intel x86-64 ISA in the K framework \cite{rosu-serbanuta-2010-jlap}, of which we have only touched a small part. Sail is a language specifically for the purpose of specifying ISAs, and it has been used to formalize parts of ARM, RISC-V, MIPS, CHERI-MIPS, IBM Power, and x86 \cite{sail-popl2019}. Our \href{https://github.com/digama0/mm0/blob/master/examples/x86.mm0}{\texttt{x86.mm0}} specification is based on a port of the Sail x86 spec. Centaur \cite{goel2020verifying} is using an x86 specification to build a provably correct chip design.

  \item Machine code verification does not differ significantly from program verification at other levels, and a number of techniques have developed to deal with it, such as Hoare logic and Separation logic. \cite{myreen2009formal} shows how machine code can be verified (in HOL4) by decompiling the machine code into HOL functions.

  \item Verified compilers are programs that produce machine code from a source language with a specified semantics, that have been proven to preserve the semantics of the input program. CompCert \cite{leroy2012compcert} is a verified compiler for a subset of C, and CakeML \cite{Kumar14} is a verified compiler for ML.

  \item For functional correctness of a particular program verified compilers are only half the story, as one must now show that the program has the correct behavior in the source language. For a language like C, this is difficult because there is no facility for doing such proofs. VST \cite{cao2018vst} is a tool for proving correctness of C programs deeply embedded as Coq terms.

  \item For languages with good source semantics, general soundness properties are useful for reducing the work of functional correctness. RustBelt \cite{jung2017rustbelt} is a project to prove soundness of the Rust type system using Iris \cite{jung2018iris}, a higher order concurrent separation logic.
\end{itemize}

\section{Conclusion}\label{sec:conclusion}
Metamath Zero is a theorem prover which is built to solve the problem of bootstrapping trust into a system. Yet at the same time it is general purpose --- it does not use a tailor-made program logic, it uses whatever axioms you give it, so it can support all common formal systems (ZFC, HOL, DTT, PA, really anything recursively enumerable). It is extremely fast, at least on hand-written inputs like \texttt{set.mm}, and can handle computer-science-sized problems.

Although the correctness theorem for \mmO\ is still ongoing, we believe there is value added in clearly delineating the necessary components for a system that pushes the boundaries of formal verification to cover as much as possible, so that we can obtain the highest level of confidence, without compromising when it comes to speed of the overall system, and without putting an upper bound on the performance of verified artifacts other than the target machine itself.

We hope to see a future where all the major theorem provers are either proven correct or can export their proofs to systems that are proven correct, so that when we verify our most important software, we bequeath the highest level of confidence we are capable of providing. It's not an impossible dream --- the technology is in our hands; we need only define the problem, and solve it.

%% Acknowledgments
\section*{Acknowledgments}
I would like to thank Norman Megill for writing Metamath, and Andr\'{e} Bacci, Wolf Lammen, David A. Wheeler, Giovanni Mascellani, Seul Baek, and Jeremy Avigad for their input and suggestions during the design phase of MM0. I thank Giovanni Mascellani for his support and for bringing my attention to the Bootstrappable project, and I thank Jeremy Avigad, Jesse Han, and Beno\^{i}t Jubin for their reviews of early versions of this work.

This work was supported in part by AFOSR grant FA9550-18-1-0120 and a grant from the Sloan Foundation.

%% Bibliography
\bibliographystyle{splncs04}
\bibliography{mm0-paper}

\end{document}